\documentclass[twocolumn,prl,amsmath,amssymb,showpacs,superscriptaddress]{revtex4-1}
\usepackage{epsf}      
\usepackage{graphicx}
\usepackage{color}
\usepackage{gensymb}

\begin{document}

\title {Synthesis and physical properties of Na$_x$TO$_2$ (T=Mn, Co) nanostructures for cathode materials in Na--ion batteries}

\author{Mahesh Chandra}
\email{These authors contributed equally to this work}
\affiliation{Department of Physics, Indian Institute of Technology Delhi, Hauz Khas, New Delhi-110016, India}
\author{Rishabh Shukla}
\email{These authors contributed equally to this work}
\affiliation{Department of Physics, Indian Institute of Technology Delhi, Hauz Khas, New Delhi-110016, India}
\author{Muhammad Rashid}
\affiliation{Department of Physics, Indian Institute of Technology Delhi, Hauz Khas, New Delhi-110016, India}
\author{Amit Gupta}
\affiliation{Department of Mechanical Engineering, Indian Institute of Technology Delhi, Hauz Khas, New Delhi-110016, India}
\author{Suddhasatwa Basu}
\affiliation{Department of Chemical Engineering, Indian Institute of Technology Delhi, Hauz Khas, New Delhi-110016, India}
\author{R. S. Dhaka}
\email{rsdhaka@physics.iitd.ac.in}
\affiliation{Department of Physics, Indian Institute of Technology Delhi, Hauz Khas, New Delhi-110016, India}

\date{\today}                                         

\begin{abstract}
We prepared Na$_x$TO$_2$ (T = Mn, Co) nanostructures to use as cathode material in Na-ion batteries. The Rietveld refinement of x-ray diffraction data confirms the hexagonal symmetry. Scanning and transmission electron microscopy measurements show hexagonal and rod-shape morphology for T = Co and Mn, respectively. The magnetic measurements indicate the presence of variable oxidation states of Mn/Co ions, but no long range ordering till 5~K. We find that the Na$_{0.6}$MnO$_2$ is highly insulating whereas Na$_{0.7}$CoO$_2$ is semiconducting in nature. The conduction in Na$_{0.7}$CoO$_2$ takes place through both variable range hopping (VRH) as well as activation mechanisms in different temperature ranges. However, for Na$_{0.6}$MnO$_2$ the VRH prevails at elevated temperatures. Furthermore, the coin cells have been fabricated and tested the electrochemical behavior using cyclic voltammetry, which confirms the reversibility of Na--ions during intercalation/de-intercalation.\\

Keywords: Na-ion battery materials; transition-metal oxides; physical properties

\end{abstract}

\maketitle

\section{\noindent ~Introduction}
In recent years, rechargeable batteries are of high demand for powering  portable electronic devices as well as electric vehicles. Therefore, searching for high performance electrode materials for Na--ion batteries has attracted great attention due to its abundance in nature and low cost raw materials for large scale energy storage applications \cite{YabuuchiNM12,WangJMCA}.  Among the materials used for electrodes (cathode/anode) in electrochemical energy storage devices, transition metal (TM) oxides are most suitable compounds owing to their layered structure and variable oxidation state of TM ions \cite{YabuuchiNM12, SuESM16,  Chemrev2013}. The layered structure allow a smooth insertion and deinsertion of Li$^+$/Na$^+$ during charging and discharging, whereas the valence of TM ion changes with addition or removal of the electrons through the external circuit. The structural stability of an electrode material is a crucial parameter as structural transformation in various layered materials has been observed during the battery cycling \cite{BerthelotNM11, XuEES17}. In this direction, the LiCoO$_2$ in pristine as well as doped form \cite{JPCREDDY2013, JPowerReddy2005}, is a well studied cathode material and extensively used in the Li--ion batteries \cite{NittaMT15, MizushimaMRB80} and similarly the Na$_x$CoO$_2$ in Na--ion batteries \cite{HanEES15}. However, considering the abundance of Mn and Fe in the Earth's crust, compounds based on these elements are particularly relevant for low cost energy storage in Na--ion batteries. On the other hand, the low cost LiMnO$_2$ is not thermodynamically stable in layered structure, but its Na counterpart, Na$_x$MnO$_2$ is stable in two structures, namely, O3 and P2 type structures \cite{HanEES15}. In these structures, the lattice is built up by sheets of edge-sharing TO$_6$ octahedra and the alkali ions are inserted between these sheets with trigonal prismatic (P) or octahedral (O) environment \cite{HanEES15}. The packing also differs in the number of sheets within the unit cell: 2 or 3, and the P2 and O3 phases have monoclinic distortions in their parent phases \cite{DelmasSSI81}. Due to a different migration pathway of Na--ion, P2-type structure gives better rate performance among these two structures \cite{SuESM16}. A small change in the Na concentration which coupled with structure causes a significant change in the Na--ion dynamics, hence the electrochemical performance of the material in the battery. For example Na$_{0.67}$MnO$_2$ possesses a 2D layered structure whereas Na$_{0.44}$MnO$_2$ has a 3D tunnel type structure \cite{MendiboureJSSC85}. Also, a small change in Na content from 0.6 to 0.7 causes a metal to insulator transition in Na$_x$CoO$_2$ \cite{KishanJAP05}, which determines the electronic transport mechanism in these materials with temperature. 

Apart from the Na stoichiometry the morphology also plays a key role in determining the capacity of the cathode materials \cite{BucherACS14, CaoAM11, XuRSC14}. Nanostructured materials have been found to improve the capacity of a Na--ion battery due to their large surface to volume ratio, which has a significant impact on diffusion of Na--ions in the solid phase. The material processing by various methods such as carbon coating, fluorination or microwave irradiation can also improve the electrochemical performance \cite{Reddy1,carboncoating}. The electrochemical potential of a cell, which determines the energy density of a battery is directly correlated with the valence state and electronic configuration of the electrode material. Therefore, in order to enhance the performance of an electrode material, an understanding of its physical ground states is vital. Despite being synthesized in 1971 by Parant {\it et al.} \cite{ParantJSSC71}, the Na$_x$MnO$_2$ family has only recently been studied for its elecrochemical properties owing to its potential as energy storage material. However, its physical properties such as electronic and magnetic ground states have largely been unexplored, which play a crucial role in determining the performance of Na--ion batteries. On the other hand, Na$_x$CoO$_2$ is well studied for its electrochemical properties as well as other physical properties such as thermoelectric, magnetic and electronic ground states, but obtaining a stable phase has always been a challenge due to its structural sensitivity towards moisture and air. The most stable structure of Na$_x$CoO$_2$ has been reported in the P2-phase (0.67$\leq$ x $\leq$ 0.72).\cite{DelmasSSI81} Here, we report the synthesis of the stable P2 phase of Na$_{x}$TO$_2$ (T = Mn,Co) nanostructures using sol-gel route and report their structural, morphological, electric transport and magnetic properties for Na--ion battery applications. In order to analyze the electrochemical behavior, cyclic voltammetry (CV) measurements were performed on fabricated coin cells and confirm the reversibility of Na$^+$ during intercalation/de-intercalation.

\section{\noindent ~Experimental}

\begin{figure}
\includegraphics[width=3.5in]{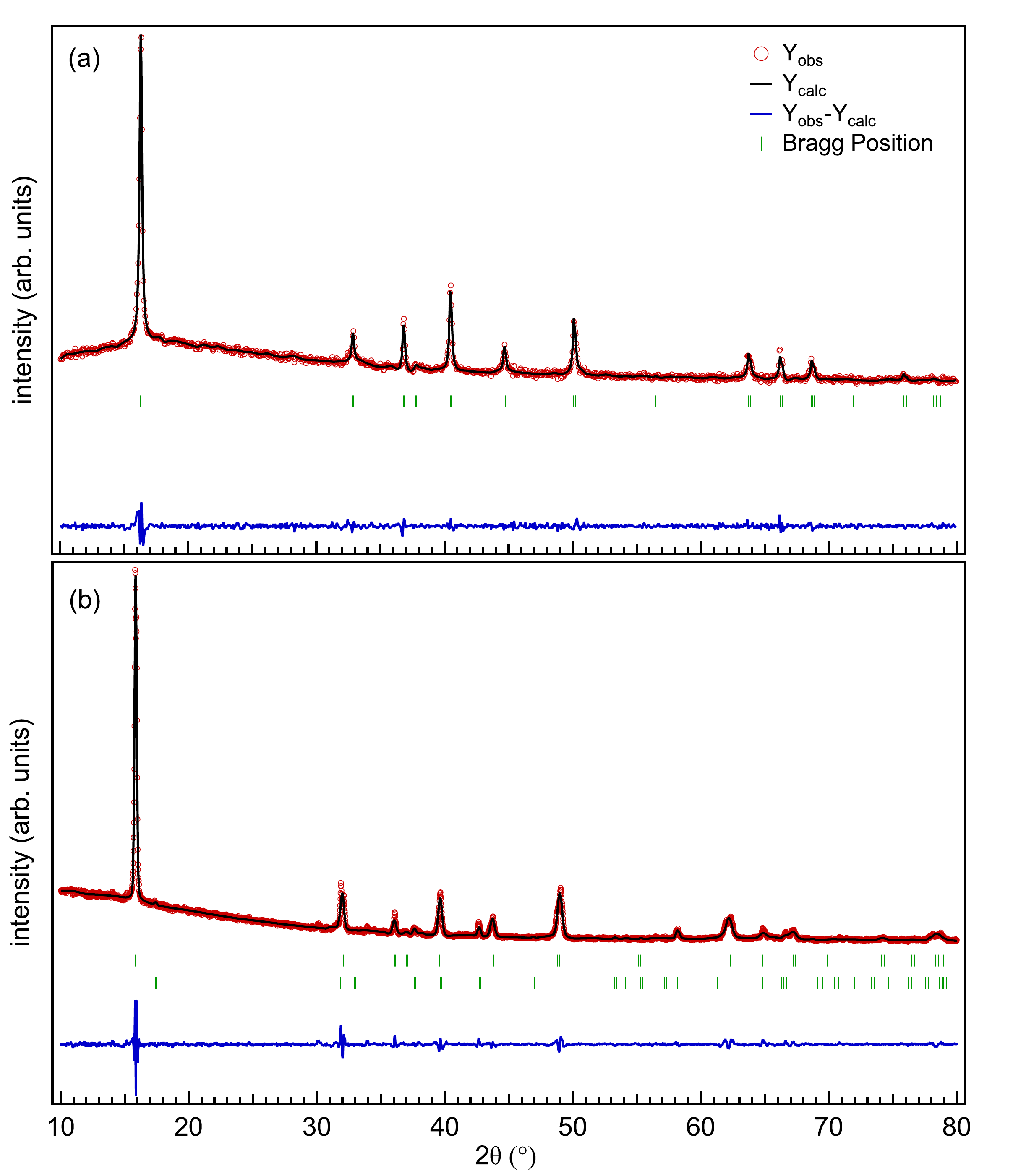}
\caption[XRD pattern with Rietveld refinement for the Na$_x$MnO$_2$ fitted to hexagonal space group ] {Room temperature XRD pattern (red) and Rietveld refinement (black) of as-prepared (a) Na$_x$CoO$_2$ and (b) Na$_x$MnO$_2$ along with Bragg peaks and residual. The refinement is performed with hexagonal P6$_3$/mmc space group (no. 194) and for Na$_x$MnO$_2$ an additional monoclinic phase has been used for better fitting.} 
\label{fig:Fig1_XRD1}
\end{figure}

We have synthesized Na$_x$TO$_2$ (T = Mn, Co) using a sol-gel method. Sodium acetate (99\%), manganese acetate (99\%), and cobalt acetate (99.9\%) from Merck were added in a stoichiometric ratio in deionized water and homogeneously mixed via stirring for 2~hrs. Then the calculated amount (molar ratio of 3:1) of citric acid (Sigma, 99.9\%), used as a complexing agent was added to the solution with overnight stirring at 90 $^0$C, which results in the formation of gel. This gel was dried at 100$^0$C for 12~hrs to low layered powder, and obtained powder then ground to fine particles. We finally sintered the powder at 700$^0$C for 10~hrs to synthesize Na$_x$CoO$_2$, whereas for Na$_x$MnO$_2$ the powder was preheated to 900$^0$C for 12~hrs and then finally heating was done at 1100$^0$C for 10~hrs. We performed the room temperature powder x-ray diffraction (XRD) with CuK$\alpha$ radiation (1.5406~\AA) from Panalytical X-ray diffractometer in the 2$\theta$ range of 10--85$^0$. We analyzed the recorded XRD data by Rietveld refinement using Fullproof package, where the background was fitted using linear interpolation between data points. The surface morphology of the prepared materials has been investigated using a low magnification field effect scanning electron microscope (FE-SEM) at 20~keV electron energy. The transmission electron microscopy (TEM) measurements have been done with JEOL JEM-1400 Plus microscope at 120~keV. Raman spectra of prepared pellets were recorded with Renishaw inviaconfocal Raman microscope at wavelength of 532~nm and grating of 2400~lines/mm with 1~mW laser power on the sample. Transport properties and temperature and field dependent magnetic measurements were done using physical property measurement system (PPMS), and SQUID from Quantum Design, USA. For the electrochemical measurements, we have assembled the coin-cells of CR2016-type, with NaTO$_2$ (T = Mn, Co) as cathode, Na thin disk as anode, glass filter as a separator, and 1.0 mole NaClO$_4$ dissolved in ethylene carbonate (EC) and propylene carbonate (PC) (in 1:1 by volume) as the electrolyte. The cell assembly is carried out in the glove-box with water and oxygen content lower than 0.5~ppm, to avoid the oxidation and reaction with moisture. A battery cycler from Bio-Logic-VMP3 has been used for the cyclic voltammetry of the fabricated cells.

\section{\noindent ~Results and discussion}

\begin{table}
		\centering
		\label{tab:rietveld}
		\caption{Rietveld refined (with P6$_3$/mmc space group) parameters from the XRD data; H is hexagonal, M is monoclinic.}
		\vskip 0.15 cm
		\begin{tabular}{|c|c|c|c|c|c|cl}
		\hline
		\textbf{Sample}&\textbf{$\chi^2$}&\textbf{R$_{wp}$(\%)}&\textbf{a (\normalfont\AA)} &\textbf{b (\normalfont\AA)}&\textbf{c (\normalfont\AA)}&\textbf{phase} \\
		\hline
		Na$_{0.7}$CoO$_2$ & 1.15 & 18.3 & 2.824 & 2.824 & 10.928& H (100\%)\\
		\hline 
		Na$_{0.6}$MnO$_2$ &2.35 & 22.8 & 2.875 & 2.875& 11.193& H (80\%)\\
	
		 & &  & 5.763 & 2.812& 5.403& M (20\%)\\
		\hline
		\end{tabular}
\end{table}

The room temperature XRD patterns of as-synthesized Na$_x$TO$_2$ (Mn, Co) samples are shown in Fig~1. For the prepared Na${_x}$CoO$_2$ sample [Fig.~1 (a)], all the peaks are well described by hexagonal P6$_3$/mmc space group and the Rietveld refined lattice parameters are in good agreement with the values reported for the composition Na$_{0.71}$CoO$_2$ (JCPDS file No. 30-1182), which is the most stable P2-phase in the all possible structures of Na$_x$CoO$_2$ \cite{DelmasSSI81,ShackletteJES88}. The XRD pattern of Na$_x$MnO$_2$ [Fig.~1 (b)] is in agreement with that of P2-type Na$_{0.6}$MnO$_2$ \cite{HernanJMC02} and the Rietveld refinement confirm the hexagonal structure with P6$_3$/mmc space group (194). However, there are few additional peaks observed in XRD pattern which corresponds to the $\alpha$-NaMnO$_2$ (monoclinic) phase. We have included this phase in the refinement  and find about 20\% contribution, which is expected as the final sintering was done at high temperature in order to stabilize the P2-phase \cite{ParantJSSC71}. 
The lattice parameter extracted from the refinement are very well in agreement with the previously reported P2-Na$_{0.6}$MnO$_2$ \cite{HernanJMC02}. We found that these samples have a stable structure when exposed to air and moisture at least up to the observed 60 days. The extracted lattice parameters for both the samples are summarized in table\ref{tab:rietveld}~I. The ratio of Na and Mn/Co has been investigated (not shown) by energy dispersive x-ray spectroscopy (EDXS), which confirm the ratio of Na:Mn as 0.57:1 in Na$_x$MnO$_2$ sample. Our structural and EDXS analysis confirm that the Na$_x$MnO$_2$ sample is stabilized in P2-Na$_{0.6}$MnO$_2$ composition.

\begin{figure}
\includegraphics[width=3.3in]{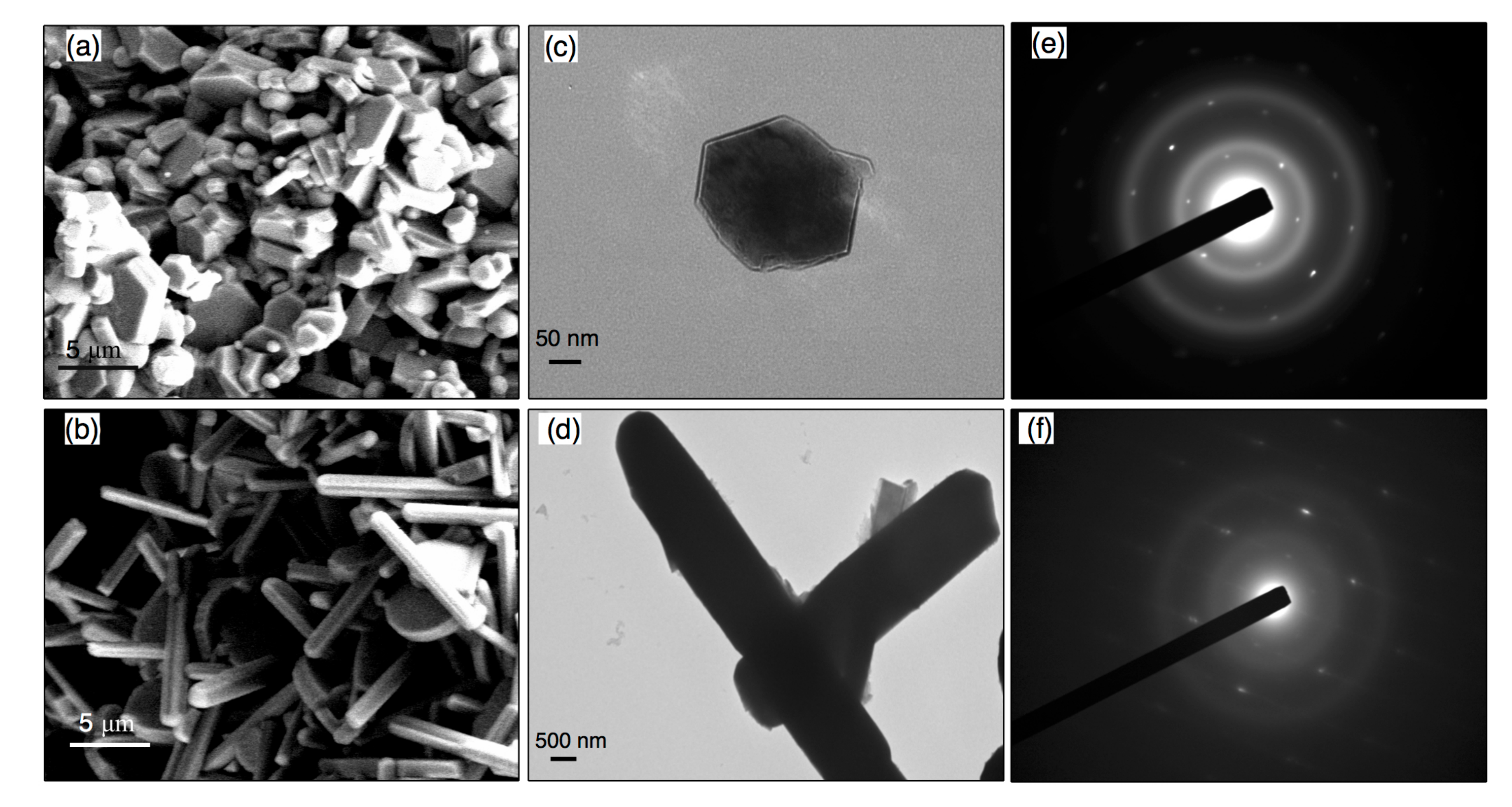}
\caption[Micrographs of prepared cathode materials via sol-gel process.]{(a-b) FE-SEM images, (c-d) TEM images, and (e-f) SAED patterns of as prepared Na$_{0.7}$CoO$_2$ and Na$_{0.6}$MnO$_2$ samples, respectively.}
\label{fig:Images}
\end{figure}

\begin{figure}
\includegraphics[width=3.6in]{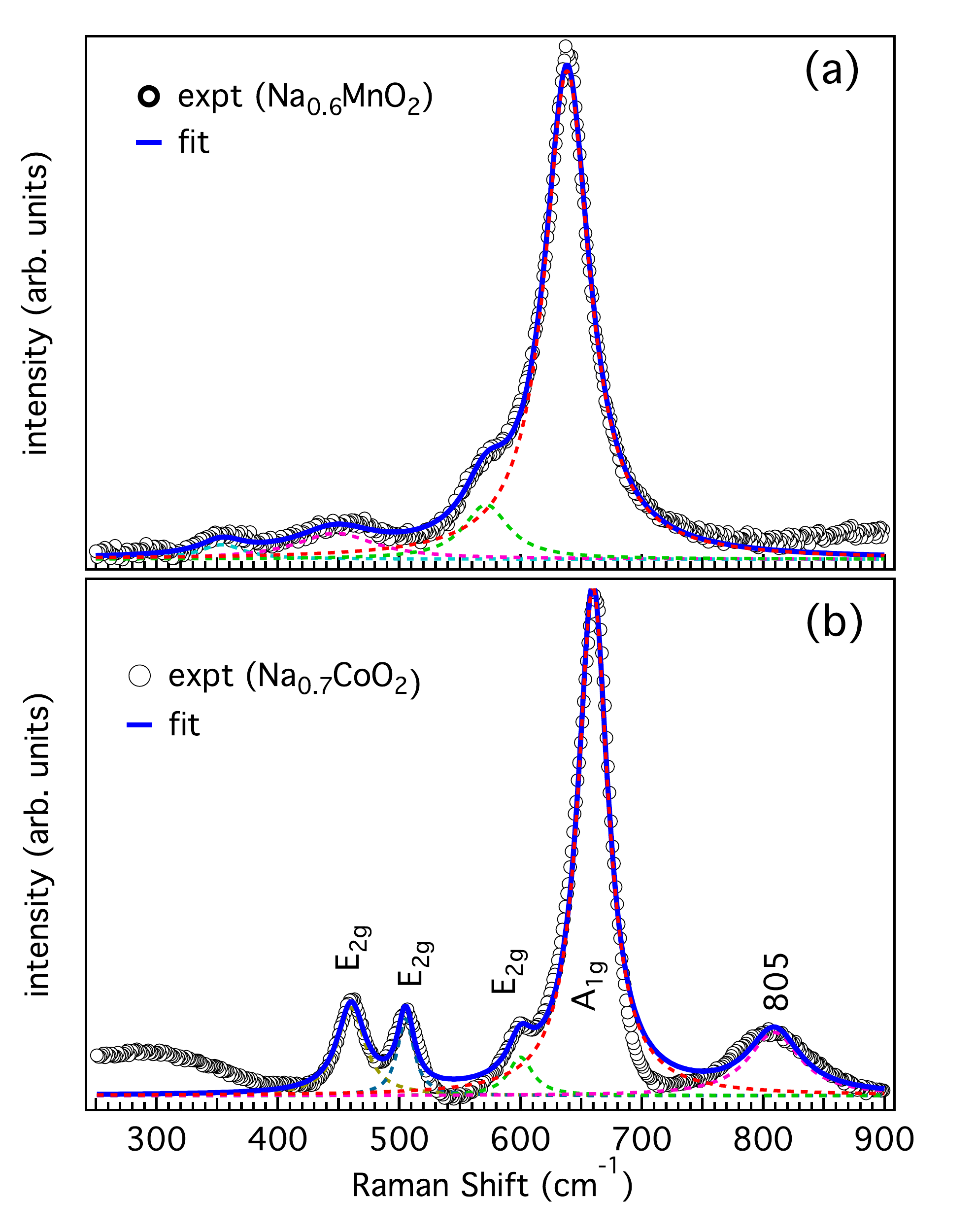}
\caption[Raman Spectra for the Na$_x$TO$_2$ pellets showing characteristic modes]{Raman spectrum of (a) Na$_{0.6}$MnO$_2$, and (b) Na$_{0.7}$CoO$_2$pellet, measured with $\lambda=$ 532~nm at room temperature, along with the fitting using Lorentzian line shape.} 
\label{fig:Raman}
\end{figure}

Figs.~\ref{fig:Images} (a--f) show the surface morphology of Na$_x$TO$_2$ (T = Mn, Co) samples, investigated using low magnification FE-SEM and TEM imaging techniques. The FE-SEM images [Figs.~\ref{fig:Images}(a, b)] show the hexagonal and rod-shaped morphology for Na$_{0.7}$CoO$_2$ and Na$_{0.6}$MnO$_2$ samples, respectively. These structures are further investigated with the TEM [Figs.~\ref{fig:Images}(c, d)], where we observed that the particle morphology in TEM is similar as in the FE-SEM, i. e., more accurately the Na$_{0.7}$CoO$_2$ in hexagonal particles of size about 200~nm, and the Na$_{0.6}$MnO$_2$ in rods of different thickness with a $\mu$m length. The selected area electron diffraction (SAED) pattern taken from these samples are shown in Figs.~\ref{fig:Images} (e, f). A hexagonal symmetry is observed in the recorded SAED patterns and the lattice constants deduced from the diffraction points (most intense) are comparable to the lattice parameter from the Rietveld refinement and (002) peak of XRD patterns. However, the SEAD for Na$_{0.7}$CoO$_2$ which is synthesized at relatively low temperature than Na$_{0.6}$MnO$_2$ shows diffused rings as well as clear bright spots. We can not conclude from this that it is nanocrystalline but such pattern can possibly be originating from  agglomerated crystals of various sizes \cite{SEAD99}. In Fig.~\ref{fig:Raman}(a), we present the Raman spectrum of the Na$_{0.6}$MnO$_2$ sample, which shows the most prominent Raman mode at 650 cm$^{-1}$ along with other modes at 580, 460, and 350 cm$^{-1}$, as shown by the de-convoluted components by dashed lines. The mode at 580 cm$^{-1}$ corresponds to vibration of Mn-O-Mn in the stretching mode and indicates the presence of Mn$^{4+}$ in the sample \cite{LiuACS16} whereas the peak at 650 cm$^{-1}$ is the characteristic of symmetric stretching vibrations of Mn-O bond \cite{LiuACS16, KarikalanCEJ17, ZhaoRSC13}. Fig. \ref{fig:Raman}(b) shows the room temperature Raman spectra for Na$_{0.7}$CoO$_2$ , where we found that the Raman active modes can be identified as A$_{1g}$ at 662 cm$^{-1}$ and E$_{2g}$ at 462, 505, and 600 cm$^{-1}$. Iliev et. al. and others reported that the A$_{1g}$ mode involve out of plane motions of only oxygen atoms in the CoO$_6$ octahedra, and E$_{2g}$ modes are connected with both Na and O motions, whereas Co motions are not Raman active \cite{IlievPhysica04, ShiPRB08, LemmensPRL06}.

\begin{figure}
\includegraphics[width=3.5in]{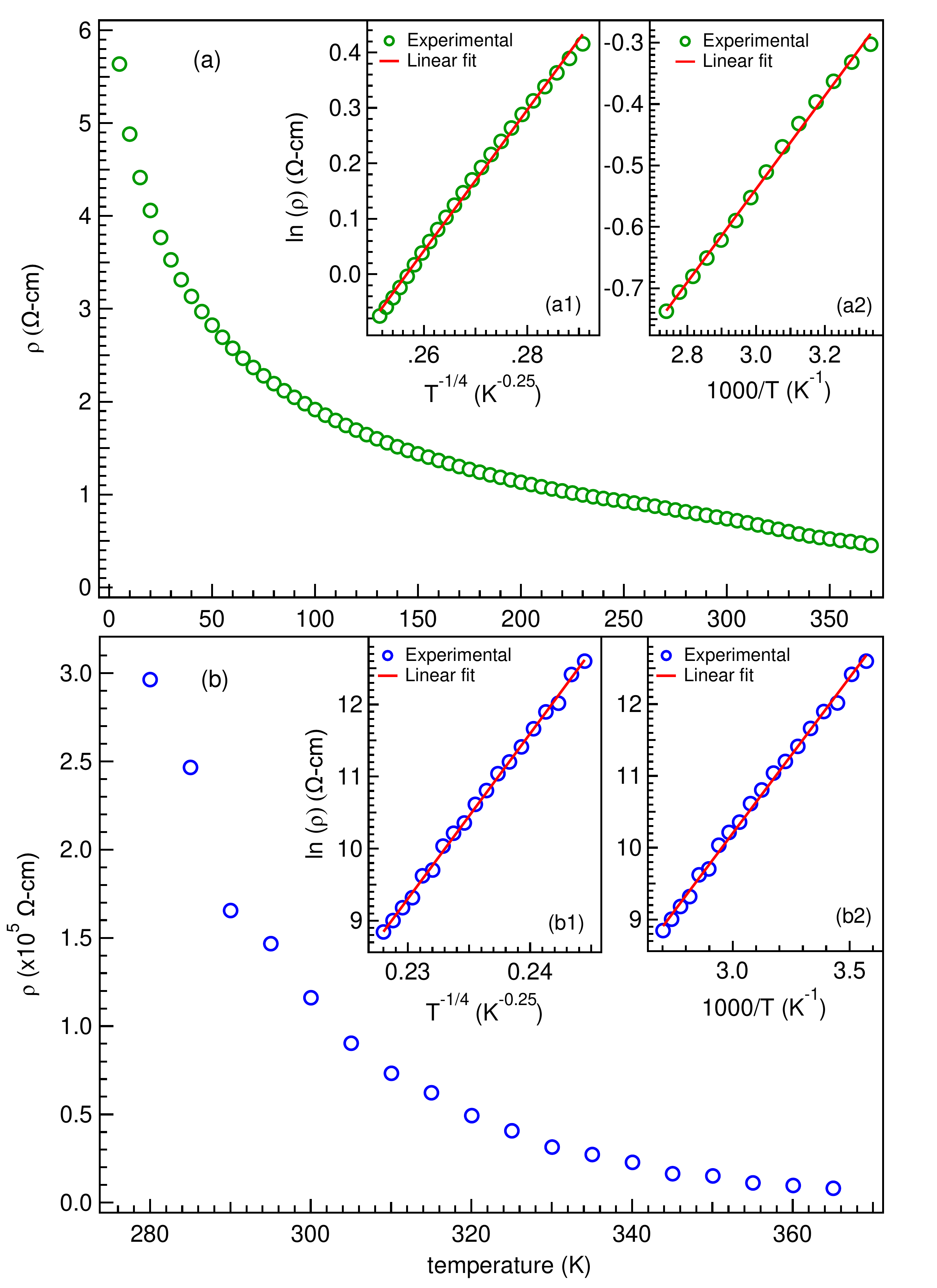}
\caption[Temperature dependent Resistance of Na$_x$MnO$_2$. The inset figure shows the data fitted to VRH model. ]{Temperature dependent resistivity of (a) Na$_{0.7}$CoO$_2$ and (b) Na$_{0.6}$MnO$_2$ samples. The inset figures show the data fitted with VRH (a1, b1) and Arrhenius (a2, b2) models.} 
\label{fig:RT}
\end{figure}

\begin{figure*}
\includegraphics[width=7in]{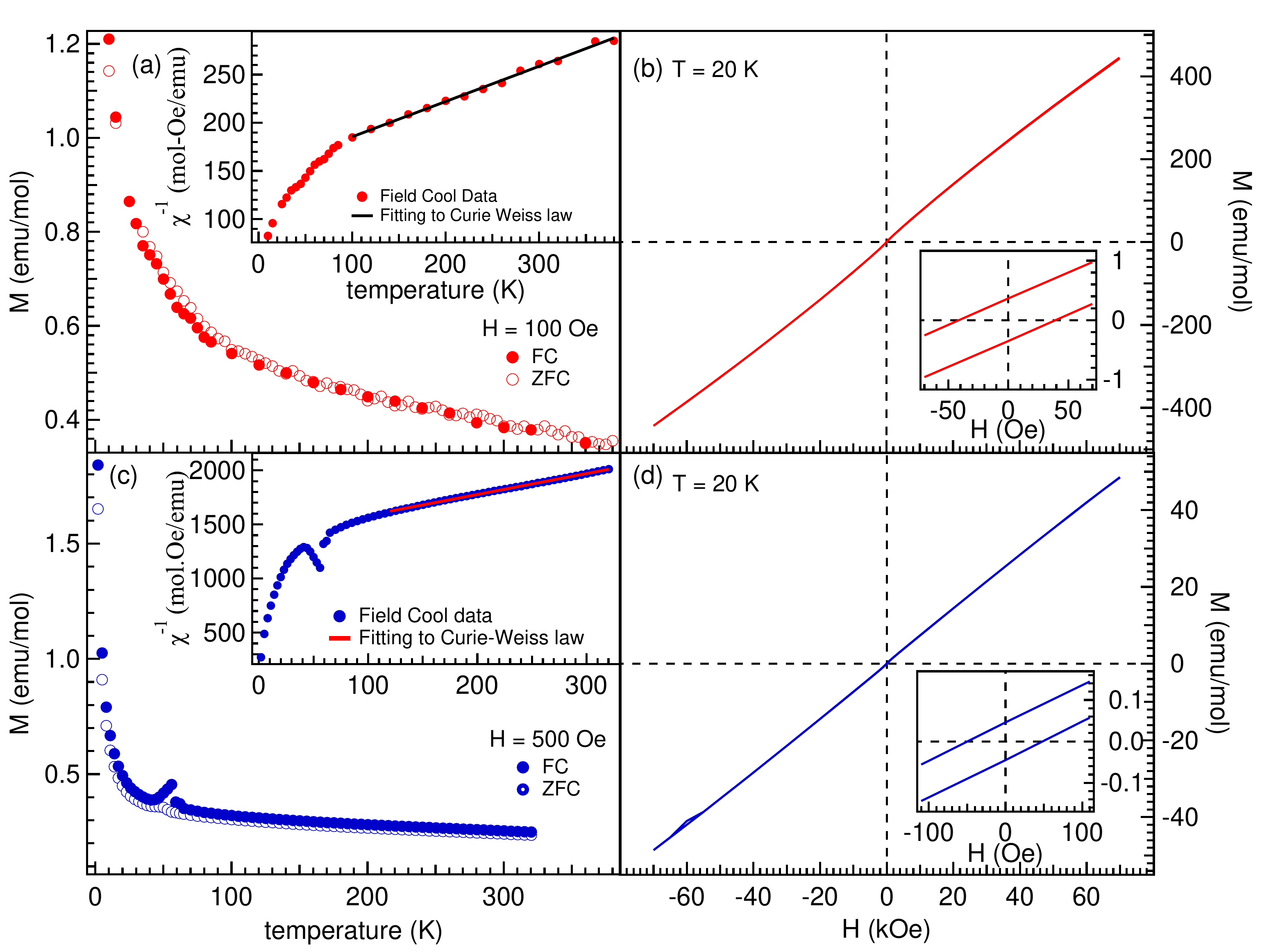}
\caption[(a)Temperature dependent magnetization of Na$_x$MnO$_2$. The inset shows the Curie-Weiss fitting till 100 K. (b)Magnetic isotherm for Na$_x$MnO$_2$ at 20 K]{The temperature dependent and isothermal magnetization of Na$_{0.6}$MnO$_2$ (a, b) and Na$_{0.7}$CoO$_2$ (c, d) samples. The insets in (a, c) show the Curie-Weiss fitting till 100-120~K. Insets in (b, d) show the zoomed view in the lower field range.}
\label{fig:M}
\end{figure*}

In order to get insight of the electronic transport properties, we have performed temperature dependent resistivity measurement from 380 K to 5~K (as shown in Fig.~\ref{fig:RT}), which show the insulating nature for both these samples throughout the respective measured temperature range. Furthermore, we performed detailed analysis to extract the physical parameters, and fitted the temperature dependent resistance with two different conduction models namely, Arrhenius model: $$\rho=\rho_0exp(E_a/k_B T)$$  where, E$_a$ is the activation energy and other one is variable range hopping (VRH) model \cite{MottTF90},
$$\rho=\rho_0exp(T_0/T)^{1/4}$$  
where, T$_0$ is the characteristic temperature. 

Here the Arrhenius model describes conduction by simple activation of charge carriers through the band gap between conduction and valance band, whereas VRH model prevails in disordered systems where potential fluctuation at spatially separated sites (due to disorder) assist the hopping of charge carriers. In VRH model, the characteristic temperature T$_0$ is related to the localization length and density of states near the Fermi level N(E) by the relation: 
$$k_BT_0=\left(\frac{18}{L^3 \times N(E)}\right)$$ 
where, L is the localization length.

\begin{table}
		\centering
		\label{tab:RT}
		\caption{Calculated values of characteristic temperature and density of states N(E) by VRH model and activation energy E$_a$ by Arrhenius model for Na$_x$TO$_2$ (T = Mn, Co).}
		\vskip 0.2 cm
		\begin{tabular}{|c|c|c|}
		\hline
		\textbf{Sample}& VRH model & Arrhenius model \\
		& T$_0$(K), N(E)(eV$^{-1}$cm$^{-3}$)& E$_a$ (meV) \\
		\hline
		Na$_{0.7}$CoO$_2$ & 3.02$\times$10$^{5}$, 9.1$\times$10$^{23}$ & 64.8$\pm$0.6 \\
		\hline 
		Na$_{0.6}$MnO$_2$ & 2.7$\times$10$^{9}$, 9.9$\times$10$^{18}$ & 374$\pm$5 \\
		\hline
		
		\end{tabular}
\end{table}

First we discuss the electronic transport in Na$_{0.7}$CoO$_2$ and present the analysis in the inset of Fig.~\ref{fig:RT} (a), where the resistivity data are fitted well with the Arrhenius model at higher temperature range 365--300~K and with the VRH model at lower temperature range of 245--165~K [see Figs.~\ref{fig:RT}(a2) and (a1), respectively]. We tried to fit the low temperature data with other models (such as adiabatic nearest neighbor hopping model of small polaron conduction and Schklovskii-Efros type of VRH with soft gap), but low temperature data of Na$_{0.7}$CoO$_2$ could not be fitted well with any of these models. Interestingly, the Na$_{0.6}$MnO$_2$ sample shows a highly insulating behavior as depicted from its high negative value of temperature coefficient of resistivity and a high resistivity at room temperature [Fig.~\ref{fig:RT} (b)]. Therefore, we could not measure the resistivity of Na$_{0.6}$MnO$_2$ sample below 280~K as the value of resistivity went past the measurement limit of the instrument. The fitting to Arrhenius model in the temperature range between 380~K to 300~K gives the activation energy 374$\pm$5 meV. However, such small band gap is expected to result in a semiconducting nature as in the case of $\alpha$ - NaMnO$_2$ (E$_a$= 450~meV) \cite{MendiboureJSSC85}. On the other hand, the data also fits well with the VRH model in the same temperature range, as shown in the inset of Fig.~\ref{fig:RT}(b) for both the models in (b2) and (b1), respectively. The extracted parameters are summarized in the table\ref{tab:RT}~II for both the samples and discussed below.

 From the fitted data, the T$_0$ is quite large for Na$_{0.6}$MnO$_2$ compare to Na$_{0.7}$CoO$_2$ indicating a small localization length and/or smaller density of states. If we take the localization length comparable to the Mn/Co-O bond length we can get an estimate of density of states [N(E)] near the Fermi level. By taking Mn-O and Co-O bond lengths 1.98 \normalfont\AA~ and 1.96 \normalfont\AA, respectively, for Na$_{0.6}$MnO$_2$ and Na$_{0.7}$CoO$_2$, we have calculated N(E), as mentioned in the table\ref{tab:RT}~II. Moreover, we found the most probable hopping distance (R) and hopping energy (W) using following equations\cite{ViretPRB97}: $$R=\left[\frac{9L}{8 \pi k_BTN(E)}\right]^{1/4}$$ and $$W=\frac{3}{{4}\pi R^3 N(E)}$$ The hopping distance (R) at 300 K come out to be 4.12$\times$10$^{-7}$ cm and hopping energy (W) is 1.25~eV for Na$_{0.6}$MnO$_2$ sample. These values fulfill the requirements of Mott VRH conduction, which are R/L$\textgreater$1 and W $\textgreater$ k$_B$T, inferring that the conduction takes place via variable range hopping in Na$_{0.6}$MnO$_2$. However, for Na$_{0.7}$CoO$_2$ the hopping energy slightly higher (1~meV) than the k$_B$T at 200~K, possibly due to the difference in estimated and actual localization length for Na$_{0.7}$CoO$_2$. We now discuss the observation of the strong electron localization in Na$_{0.6}$MnO$_2$, and there can be two plausible reasons for such strong localization: first the presence of Mn$^{3+}$ and Mn$^{4+}$ where only Mn$^{3+}$ is strong Jahn-Teller active, which makes the system strongly disorder and hence highly insulating \cite{MendiboureJSSC85}. As discussed above, the finding that the VRH conduction is taking place in this sample support this possibility. Also, a cooperative Jahn-Teller distortion coupled with Na vacancy ordering and charge ordering of Mn$^{3+}$ and Mn$^{4+}$ have recently been observed in Na$_{5/8}$MnO$_2$ \cite{LiNM14}. This suggest the second possibility, which is the charge ordering of Mn$^{3+}$ and Mn$^{4+}$. However, charge ordering generally takes place at lower temperatures in TM oxides and the temperature range of fitting is very high. Therefore, further investigation is required to pin point exact reason for the highly insulating behavior observed in Na$_{0.6}$MnO$_2$.

We now move to the discussion of magnetic behavior of these Na$_x$TO$_2$ (T = Mn, Co). The magnetic ground state in Na$_x$TO$_2$ is a result of super exchange interaction between neighboring T ions and whether the interaction is antiferromagnetic (AFM) or ferromagnetic (FM) depends upon the valence states of these ions. The valence state not only determines the redox potential but also the structural stability of the electrode. For example in the case of T=Mn, the higher concentration of Mn$^{3+}$ means higher Jhan teller distortion and thus a poor cycling performance \cite{HernanJMC02}. Recent study shows that for an optimum insertion of Na/Li/K ion in MnO$_2$ layered structure give rise to a FM ground in an otherwise AFM lattice \cite{TsengNATSCI15}. The study indicates that the storage of these alkali metals and magnetic ground states of host are correlated. 

In order to get the insight of the magnetic ground state of Na$_{0.6}$MnO$_2$, which is largely unexplored for this composition \cite{AbakumovCM14}, we first performed temperature dependent magnetization (M--T) as shown in Fig.~\ref{fig:M} (a, c). The zero field cooled (ZFC) and field cooled (FC) curves do not show any transition till 5~K indicating the absence of long range magnetic ordering in both the samples. However, the inverse susceptibility ($\chi^{-1}$) versus temperature plots [insets of Figs.~\ref{fig:M} (a, c)] deviate from the Curie-Weiss behavior below 100-120~K. The susceptibility obeys the Curie-Weiss law at high temperatures from 100--120~K to 380~K. The Curie constant (C) and $\theta$ comes out as 2.71 and -403~K, respectively from the fitting for Na$_{0.6}$MnO$_2$ sample. Also, the effective magnetic moment ($\mu_{eff}$) has been calculated from the Curie constant, which found to be 4.65~$\mu_B$. By taking the proportion of Mn$^{3+}$ (high spin) and Mn$^{4+}$ in accordance with Na$_{0.6}$MnO$_2$, the theoretically calculated value of $\mu_{eff}$  is about 4.5~$\mu_B$, which is very close to the experimentally observed value. 

 \begin{figure}
\includegraphics[width=3.6in]{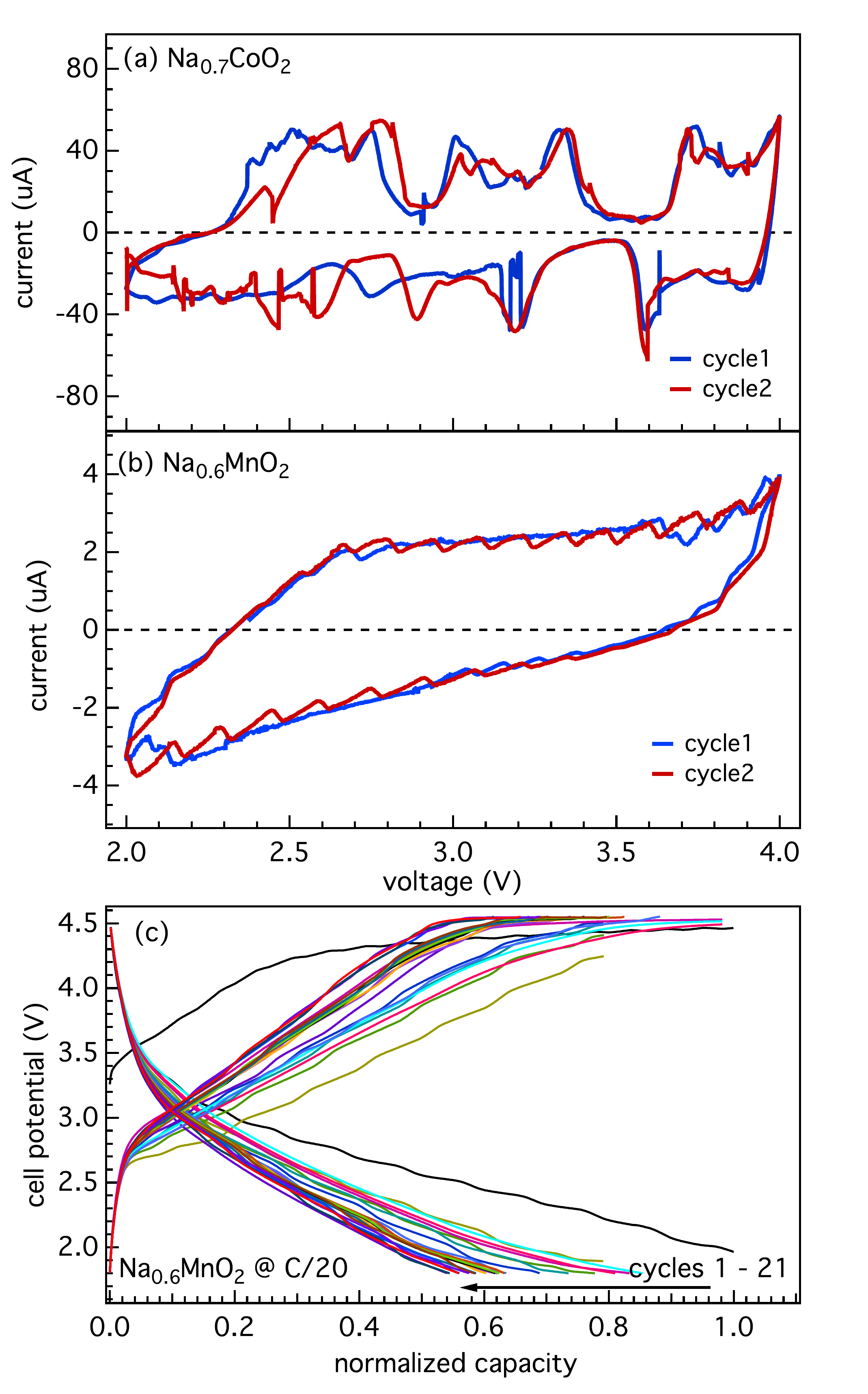}
\caption {The cyclic voltammetry (CV) characterization of fabricated coin cells of (a) Na$_{0.7}$CoO$_2$ and (b) Na$_{0.6}$MnO$_2$ with NaClO$_4$ as the electrolyte and metallic Na as the counter electrode, measured between 2.0--4.0~V at a scan rate of 0.1~mV/s. (c) The normalized capacity of Na$_{0.6}$MnO$_2$ coin-cell.}
\label{fig:CV}
\end{figure}

In order to confirm the absence of long range magnetic ordering we performed magnetic field dependent magnetization at 20~K as shown in Figs.~5(b, d). The magnetic isotherm plots show a typical paramagnetic behavior for both the samples with a very small hysteresis of 40--50~Oe. Similar magnetic behavior has been reported in Co substituted NaMnO$_2$ samples \cite{CarlierDT11}. It is quite interesting that despite of a possible antiferromagneic interaction between Mn$^{3+}$- Mn$^{3+}$ and ferrimagnetic between Mn$^{3+}$ and Mn$^{4+}$, this sample shows a paramagnetic behavior \cite{LiNM14}.  In the case of Na$_{0.7}$CoO$_2$, a small peak around 60~K is observed in the M--T data [Fig.~5(c)], which could be due to small contamination of oxygen at the sample surface as the SQUID is highly sensitive. For Na$_{0.7}$CoO$_2$, we found that Curie-Weiss constants C and $\theta$ are 0.52 and -719 K, which gives the effective magnetic moment of about 2.1~$\mu_B$. As the Na concentration is lower than one, Co ions present in both 3+ and 4+ oxidation states and can exist in low spin (LS), intermediate state (IS), high spin (HS) or combination of mixed states. Therefore, by taking Co$^{4+}$ in LS and Co$^{3+}$ in mixed of LS and IS states (30\% and 70\%), we have calculated the theoretical magnetic moment, which turned out to be the calculated value of $\mu_{eff}$ = 2.1. Our magnetization study confirm the variable oxidation states of TM ions, which are important for the performance of cathode materials.  

Finally, we discuss electrochemical characterizations of prepared coin cells using Na$_{0.7}$CoO$_2$ and Na$_{0.6}$MnO$_2$ as working electrode against Na counter electrode. Figs 6(a, b) show the cyclic voltammograms (CV) of both the electrodes for first few cycles, measured between 2.0--4.0 V at a scan rate of 0.1 mV/s at 20$^0$C. In the CV curves of  Na$_{0.7}$CoO$_2$, we observed several peaks during anodic as well as the corresponding cathodic scans which is in agreement with the literature \cite{BerthelotNM11,ReddyJMCA15,BhidePCCP14,SauvageIC07}. This confirms the reversibility of Na--ions and indicates the presence of various intermediate valence states during intercalation/de-intercalation of Na--ion in the solid matrix of the Na$_{0.7}$CoO$_2$. Similarly in the case of Na$_{0.6}$MnO$_2$ a multiphase evolution is observed during cycling as depicted by the presence of manifold oxidation and reduction peaks, which is common feature for Na$_x$MnO$_2$ \cite{HernanJMC02, MaJECS11, BucherJSSE2013}. The galvanostatic charging-discharging profile for Na$_{0.6}$MnO$_2$ is shown in Fig.~6(c), for first 21 cycles at current rate of 8 mA/g. We show the normalized capacity in Fig.~6(c) since the measured charge/discharge profiles of Na$_{0.6}$MnO$_2$ coin cells exhibit capacity that is lower than the theoretical value. The curves shows plateaus consistent with the multiphase evolution seen in CV. Each plateau correspond to extraction of Na ion from different sites. The capacity degradation upon cycling is fast and is in agreement with the earlier report \cite{HernanJMC02} which is owing to the presence of Jahn-Teller Mn$^{3+}$ ion. However, the durability for T = Co is better \cite {ShackletteJES88} due to the absence of Jahn-Teller distortion. 

\section*{Conclusions}
In conclusion, we have synthesized a stable hexagonal phase of Na$_x$TO$_2$ (T = Mn, Co) and report the structure, morphology, magnetic ordering and electronic ground state, which play crucial role in determining the electrochemical properties of these compounds in Na-ion battery applications. The Na$_{0.6}$MnO$_2$ sample is highly insulating and the conduction takes place by variable range hopping of charge carriers even at higher temperatures. One the other hand Na$_{0.7}$CoO$_2$ shows semiconducting nature and conduct via activation and variable range hooping at high and low temperatures, respectively.  Despite of the presence of magnetic Mn/Co$^{3+}$ and Mn/Co$^{4+}$, a long range magnetic ordering is absent in these nanostructured samples till low temperatures. The cyclic voltammogram curves of the electrodes indicate the reversibility of Na--ions during intercalation/de-intercalation. The capacity degradation is fast in Na$_{0.6}$MnO$_2$ possibly due to structural distortion upon cycling and the presence of Jahn-Teller Mn$^{3+}$ ion compare to Na$_{0.7}$CoO$_2$. 

\section*{\noindent ~Acknowledgments}
We acknowledge the financial support from IIT Delhi through the FIRP project (no. MI01418). MC, and RS thank SERB-DST (NPDF, no PDF/2016/003565), and MHRD, respectively, for the fellowship. MC and MR also thank IIT Delhi for postdoctoral fellowship through FIRP project. Priyanka is thanked for useful discussions. Authors acknowledge IIT Delhi for providing research facilities: XRD, Raman, �PPMS EVERCOOL-II� at physics department, and EDX, SEM, TEM at CRF. We also thank the physics department, IIT Delhi for support. RSD thank SERB-DST for support through Early Career Research (ECR) Award (project reference no. ECR/2015/000159) and BRNS for support through DAE Young Scientist Research Award (project sanction no. 34/20/12/2015/BRNS).

\end{document}